\documentclass[11pt,a4paper,english,amssymb,nofootinbib,superscriptaddress,showpacs]{revtex4}
\usepackage{lmodern}

\usepackage[T1]{fontenc}
\usepackage[latin9]{inputenc}
\usepackage{babel}
\usepackage{amsmath}
\usepackage{graphicx}
\usepackage{amssymb}
\usepackage{esint}
\usepackage{color}
\usepackage{soul}

\usepackage[unicode=true, pdfusetitle,
bookmarks=true,bookmarksnumbered=false,bookmarksopen=false,
breaklinks=false,pdfborder={0 0 1},backref=false,colorlinks=false]
{hyperref}
\def\b{\begin{equation}}
\def\e{\end{equation}}

\makeatletter
\@ifundefined{textcolor}{}
{%
 \definecolor{BLACK}{gray}{0}
 \definecolor{WHITE}{gray}{1}
 \definecolor{RED}{rgb}{1,0,0}
 \definecolor{GREEN}{rgb}{0,1,0}
 \definecolor{BLUE}{rgb}{0,0,1}
 \definecolor{CYAN}{cmyk}{1,0,0,0}
 \definecolor{MAGENTA}{cmyk}{0,1,0,0}
 \definecolor{YELLOW}{cmyk}{0,0,1,0}
 }

\usepackage{latexsym}\usepackage{bm}

\makeatother

\makeatother

\begin{document}
\title{ Weak Field Limit of Infinite Derivative Gravity}
\author{Ercan Kilicarslan}

\email{ercan.kilicarslan@usak.edu.tr}

\affiliation{Department of Physics,\\
 Usak University, 64200, Usak, Turkey}

\begin{abstract}
A form of infinite derivative gravity is free from ghost-like instabilities with improved small scale behavior. In this theory, we calculate the tree-level scattering amplitude and the corresponding weak field potential energy between two localized covariantly conserved spinning point-like sources that also have velocities and orbital motion. We show that the spin-spin and spin-orbit interactions take the same form as in Einstein's gravity at large separations, whereas at small separations, the results are different. We find that not only the usual Newtonian potential energy but also the spin-spin and spin-orbit interaction terms in the potential energy are non-singular as one approaches $r\to 0$. 
\end{abstract}

\maketitle

\section{Introduction}
Although General Relativity (GR) provides very successful solutions, observations and predictions at the intermediate regimes, it fails to be a complete theory at both large (IR) and small (UV) scales. In the IR regime, GR does not give explanations to the accelerating expansion of the universe and rotational speed of galaxies without assuming a tremendous amount of dark energy and dark matter compared to the ordinary matter. As for small distances at the quantum level, it is a non-renormalizable theory according to perturbative quantum field theory perspective because of the infinities appearing in a renormalization procedure. These infinities coming from the self-interactions of gravitons (in the pure gravity case) cannot be regulated by a redefinition of finite numbers of parameters. GR has also black hole or cosmological type singularities at the classical level. The GR is expected to be modified at both regimes in order to have a complete theory. Here, the main question is what kind of modification in the UV will provide a complete model which may also solve cosmological or black hole singularity problems. In this respect, a possible way out of this problem was to add scalar higher order curvature terms to Einstein's theory such as the quadratic theory
\begin{equation}
I=\int d^4x(\sigma R+\alpha R^2+\beta R_{\mu\nu}^2),
\end{equation}
which describes massive and massless spin-$2$ gravitons together with a massless spin-$0$ particle \cite{stelle}. By adding higher curvature terms, renormalizability is gained, but the unitarity (ghost and tachyon-free) of the theory is lost due to a conflict between the massless and massive spin-$2$ excitations. In other words, the theory has Ostragradsky type instabilities at the classical level which become ghosts at the quantum theory. Theory has an unbounded Hamiltonian density from below. That is to say, the addition of higher powers of curvature causes a conflict between the unitarity and the renormalizability. 

On the other hand, it has been recently demonstrated that infinite derivative gravity (IDG) \cite{Biswas1,Biswas2} has the potential to have a complete theory in the UV scale\footnote{For recent developments on IDG, see \cite{Tomboulis,Biswas4, Modesto,Biswas5,Biswas6,Buoninfante:2018xiw,Modesto1,Modesto2,Modesto22,Modesto3, Talaganis:2016ovm,Edholm:2016hbt,Modesto4,Boos:2018bxf,Moffat1,Moffat2,Briscese:2012ys}.}. IDG is described by an action constructed from non-local functions $F_i(\Box)$ [given in Eq.(\ref{idfunc})], where $\Box$ is the d'Alembartian operator ($\Box=g^{\mu\nu}\nabla_\mu\nabla_\nu$). The propagator of the IDG in a flat background in $3+1$ dimensions 
\begin{equation}
\Pi_{IDG}=\frac{P^2}{a(k^2)}-\frac{P_s^0}{2a(k^2)}=\frac{\Pi_{GR}}{a(k^2)},
\end{equation}
is given in terms of Barnes-Rivers spin projection operators ($P^2,P_s^0$) \cite{Biswas1}. Here $a$ is given in terms of $F_i(\Box)$ [see Eq.(\ref{relations})] and $\Pi_{GR}$ is the pure GR graviton propagator. One of the important points is to avoid introducing ghost-like instabilities and having additional scalar degrees of freedom other than the massless spin-$2$ graviton. To do this, $a(k^2)$ can be chosen to be an exponential of an entire function as $a(k^2)=e^{\gamma(\frac{k^2}{M^2})}$, where $\gamma(\frac{k^2}{M^2})$ is an entire function. This choice guarantees that the propagator has no additional poles other than massless graviton, in other words, $a(k^2)$ has no roots. In the $a(k^2)\to 0$ or $k\ll M$ limit, the propagator takes the usual Einsteinian form. Furthermore, as the propagator does not have any extra degrees of freedom, the modified propagator is free from ghost-like instabilities. The Hamiltonian density is bounded from below. Moreover, in \cite{Talaganis}, it has been recently shown that loop divergences beyond one-loop may be handled by introducing some form factors. Furthermore, infinite derivative extension of GR may resolve the problem of singularities in black holes and cosmology \cite{Biswas1, Biswas2,Tomboulis,Biswas4, Modesto,Biswas5,Biswas6,Buoninfante:2018xiw}.  

In this work, we would like to explore the weak field limit of the IDG and compare it with the result of GR. In \cite{Biswas1}, the Newtonian potential for the point source was calculated for the IDG, here we extend this discussion to include the spin, velocities and orbital motion of the sources. By spin, we mean the rotation of the sources about their own axes. Therefore we calculate the spin-spin and spin-orbit interactions between two massive sources in IDG and show that the mass-mass interaction, the spin-spin interaction and the spin-orbit interaction part become non-singular as $r\to 0$. These non-singular results in IDG show that the theory has improved behavior in the small scale compared to GR. 

The layout of the paper is as follows: In Sec. II, we investigate the spin-spin interactions of localized point-like spinning massive objects in IDG and consider the large and small distance limits of potential energy. Section III is devoted to extend the calculations in the previous section to the case that the massive spinning  sources are also moving. In that section, in addition to mass-mass and spin-spin interactions, we studied the spin-orbit interactions in IDG. In conclusions and further discussions, we give the final result for a gravitational memory effect in IDG and discuss the effects of mass scale of non-locality on memory. In the Appendix, we give some of the details of calculations for Sec. III. 

\section{Scattering Amplitude in IDG}
The matter coupled Lagrangian density of IDG is \cite{Biswas1}
\begin{equation}
\mathcal{L}= \sqrt{-g}\bigg[\frac{M^2_P}{2}  R\ +\frac{1}{2} R F_1 (\Box) R + \frac{1}{2} R_{\mu\nu} F_2(\Box) R^{\mu\nu}
       + \frac{1}{2}C_{\mu\nu\rho\sigma} F_3(\Box) C^{\mu\nu\rho\sigma}+\mathcal{L}_{matter}\bigg],
\end{equation}
where $M_P$ is the Planck mass, $R$ is the scalar curvature, $R_{\mu\nu}$ is the Ricci tensor and $C_{\mu\nu\rho\sigma}$ is the Weyl tensor. The infinite derivative functions $F_i(\Box)$ are given as
\begin{equation}
F_i(\Box)=\sum_{n=1}^{\infty}f_{i_n}\frac{\Box^n}{M^{2n}}, 
\label{idfunc}
\end{equation} 
which are functions of the d'Alembartian operator. Here, $f_{i_n}$ are dimensionless coefficients and $M$ is the mass scale of non-locality. The linearized field equations around a Minkowski background of $g_{\mu\nu}=\eta_{\mu\nu}+h_{\mu\nu}$ reads\footnote{We will work with the mostly plus signature $\eta_{\mu\nu}={\it{diag}}(-1,1,1,1)$.} \cite{Biswas1}
\begin{equation}
a(\Box)R^L_{\mu\nu}-\frac{1}{2}\eta_{\mu\nu}c(\Box)R^L-\frac{1}{2}f(\Box)\partial_\mu\partial_\nu R^L=\kappa T_{\mu\nu},
\label{FeqIDG}
\end{equation}
where $L$ refers to linearization and non-linear functions are defined as
\begin{equation}
\begin{aligned}
 &a(\Box) =1 + M^{-2}_P \left(F_2(\Box) 
        + 2  F_3(\Box)\right) \Box, \\
&        c(\Box) = 1 - M^{-2}_P\left(4 F_1(\Box) +  F_2(\Box)  
        - \frac{2}{3}F_3(\Box)\right)\Box,\\
&        f(\Box) =M^{-2}_P \left(4F_1(\Box) + 2F_2(\Box) +\frac{4}{3} F_3(\Box)\right),
        \end{aligned}
        \label{relations}
\end{equation}
which give the constraint $a(\Box)-c(\Box) = f(\Box)\Box$. After plugging the relevant linearized curvature tensors \cite{deser_tekin} into (\ref{FeqIDG}), one arrives at the linearized field equations
\begin{equation}
\begin{aligned}
  &\frac{1}{2} \bigg[a(\Box)\left(\Box h_{\mu\nu} 
        -\partial_\sigma
        \left(\partial_\mu h^\sigma_\nu + \partial_\nu h^\sigma_\mu \right)\right)
        + c(\Box) \left(\partial_\mu \partial_\nu h + \eta_{\mu\nu} \partial_\sigma \partial_\rho 
        h^{\sigma\rho}-\eta_{\mu\nu} \Box h\right)\\&+ f(\Box) \partial_\mu \partial_\nu 
 \partial_\sigma \partial_\rho 
h^{\sigma\rho}\bigg]=-\kappa T_{\mu\nu} .
\label{IDGfeq1}
\end{aligned}
\end{equation}
 If we set $a(\Box)=c(\Box)$, we recover the pure GR propagator in the large distance limit without introducing additional degrees of freedom. Then, in the de Donder gauge $\partial_\mu h^{\mu\nu} = \frac{1}{2} \partial^\nu h$, the linearized field equations (\ref{IDGfeq1}) take the following compact form
\begin{equation}
  a(\Box){\cal{G}}^L_{\mu\nu}=\kappa T_{\mu\nu} , 
  \label{IDGfeq2}
\end{equation}
where ${\cal{G}}^L_{\mu\nu}$ is the linearized Einstein tensor defined as ${\cal{G}}^L_{\mu\nu}=-\frac{1}{2}(\Box h_{\mu\nu}-\frac{1}{2}\eta_{\mu\nu}\Box h)$. Manipulation of (\ref{IDGfeq2}) yields
 \begin{equation}
  a(\Box) \Box h_{\mu\nu}=-2\kappa(T_{\mu\nu}-\frac{1}{2}\eta_{\mu\nu}T) ,
  \label{IDGfeq3}
\end{equation}
which is the equation that we shall work with.

From now on, we consider the tree-level scattering amplitude between two spinning conserved point-like sources and find the corresponding weak field potential energy. To do that, one needs to first eliminate the non-physical degrees of freedom from the theory. For this purpose, let us consider the following decomposition of the spin-$2$ field
\begin{equation}
  h_{\mu\nu} \equiv  h^{TT}_{\mu\nu}+\bar{\nabla}_{(\mu}V_{\nu)}+\bar{\nabla}_\mu \bar{\nabla}_\nu \phi+\bar{g}_{\mu\nu} \psi,
  \label{dectth}
 \end{equation}
 where $ h^{TT}_{\mu\nu} $ is the transverse-traceless part of the field, $ V_\mu $ is the transverse helicity-$1$ mode and $ \phi $ and $ \psi $ are scalar helicity-$0$ components of the field. To obtain  $\psi $ in terms of  field $h$, one needs to take the trace  and double divergence of (\ref{dectth}) to arrive at
\begin{equation}
 h=\partial^2\phi+4 \psi, \hskip 1 cm \frac{1}{2}\partial^2h=\partial^4\phi+\partial^2\psi,
 \label{scaleinsstre}
\end{equation}
where we used $\partial^\mu\partial^\nu h_{\mu\nu}=\frac{1}{2}\partial^2h$. Then, by using (\ref{scaleinsstre}) and (\ref{IDGfeq2}), one obtains
 \begin{equation}
  \psi = \frac{\kappa}{3}(a(\Box)\partial^2)^{-1}T.
  \label{psittt}
 \end{equation}
On the other hand, inserting (\ref{dectth}) into (\ref{IDGfeq2}) yields the wave-type equation
 \begin{equation}
    h^{TT}_{\rho\nu}= -2\kappa\,{\cal O}^{-1} T^{TT}_{\rho\nu},
 \label{htt}
 \end{equation}
where the corresponding scalar Green's function is
\begin{equation}
  G ({\bf x},{\bf x}^{'},t,t^{'})={\cal O}^{-1} \equiv (a(\Box)\partial^2)^{-1}.
\end{equation}
Accordingly, the tensor decomposition of energy momentum tensor  $T_{\rho\nu} $ can be given as \cite{Gullu:2009vy}
\begin{equation}
 T^{TT}_{\rho\nu}=T_{\rho\nu}-\frac{1}{3} \bar{g}_{\rho\nu}T+\frac{1}{3} \Big (\bar{\nabla}_\rho \bar{\nabla}_\nu  \Big ) \times (\bar{\square} )^{-1} T.
\label{ttstrener}
 \end{equation}
Recall that the tree-level scattering amplitude between two  sources via one graviton exchange is given by
\begin{equation}
\begin{aligned}
 {\cal A}&=\frac{1}{4} \int d^4 x \sqrt{-\bar{g}} T^{'}_{\rho\nu}(x) h^{\rho\nu}(x) \\
 &=\frac{1}{4} \int d^4 x \sqrt{-\bar{g}} (T^{'}_{\rho\nu} h^{TT\rho\nu}+T^{'} \psi ).
 \label{scatdef}
\end{aligned}
 \end{equation}
Consequently, by plugging (\ref{psittt}),(\ref{htt}) and (\ref{ttstrener}) into (\ref{scatdef}), the scattering amplitude in a flat background can be obtained as follows
\begin{equation}
 \begin{aligned}
  4{\cal A}=-2\kappa T^{'}_{\rho\nu} {\cal O}^{-1}  T^{\rho\nu}+\kappa T^{'}{\cal O}^{-1}T,
\label{mainressct}
\end{aligned}
\end{equation}
where the integral signs are suppressed for notational simplicity. Now, we are ready to compute the tree-level scattering amplitude for IDG  between two covariantly conserved point-like spinning sources. For this purpose, let us consider the following localized spinning energy-momentum tensors 
\begin{equation}
T_{00}= m_a \delta^{(3)}({\bf x}-{\bf x}_a),\qquad T^{i}{_0}=-\frac{1}{2} J^k_a \epsilon^{ikj} \partial_j \delta^{(3)}({\bf x}-{\bf x}_a),
\label{sources}
\end{equation}
where $m_a $ are the mass and $ J_a $ are the spin of the sources which have no dimension in our limits; here $a=1,2$.
In this respect,  we want to solve the linearized IDG equations for the sources given in (\ref{sources}). The scattering amplitude (\ref{mainressct}) can be explicitly recast as
\begin{equation}
4A 
  =-2\kappa T{}_{00}^{\prime}\left\{ \frac{1}{a(\Box)\partial^2}\right\} T^{00}
+\kappa T^{\prime}\left\{ \frac{1}{a(\Box)\partial^2}\right\} T 
+4\kappa T{}_{0i}^{\prime}\left\{ \frac{1}{a(\Box)\partial^2}\right\} T^i\,_{0}.
\label{scatflat2}
\end{equation}
On the other hand one must keep in mind that, to avoid ghosts, $a(\Box)$ must be an entire function. For simplicity, let us choose  $a(\Box)=e^{-\frac{\Box}{M^2}}$ with which the main propagator can be computed as 
\begin{equation}
G({\bf x}, {\bf x}^{'}, t, t^{'})=\frac{1}{4\pi r} \mbox{erf} (\frac{Mr}{2})\delta({\bf x}-{\bf x}'-(t-t')),
\label{decgreen1}
  \end{equation}
where $ r = \lvert {\bf x}_1-{\bf x}_2 \rvert$ and $\mbox{erf} (r)$ is the error function defined by the integral
\begin{equation}
\mbox{erf} (r)=\frac{2}{\sqrt{\pi}}\int_0^r e^{-k^2}dk.
\end{equation}
Thus, by substituting (\ref{decgreen1}) into (\ref{scatflat2}) and carrying out the time integrals, one gets
\begin{equation}
 \begin{aligned}
4\,{\cal U}=&-2\kappa m_1 m_2\int d^3 x \int d^3 x^{'}\,\,\,\delta^{(3)}({\bf x}^{'}-{\bf x}_2)  \hat{G}({\bf x}, {\bf x}^{'}) \delta^{(3)}({\bf x}-{\bf x}_1)\\& +  \kappa m_1 m_2 \int d^3 x \int d^3 x^{'}\,\,\, \delta^{(3)}({\bf x}^{'}-{\bf x}_2) \hat{G}({\bf x}, {\bf x}^{'}) \delta^{(3)}({\bf x}-{\bf x}_1) \\&
+\kappa\int d^3 x \int d^3 x^{'}\,\,J_1^k\,\epsilon^{ikj} \partial'_j \delta^{(3)}({\bf x}^{'}-{\bf x}_2)\hat{G}({\bf x}, {\bf x}^{'}) \,J_2^l\,\epsilon^{ilm}\partial_m\delta^{(3)}({\bf x}-{\bf x}_1).
  \end{aligned}
\end{equation}
Here, the potential energy is $ {\cal U}={\cal A}/t $ \cite{Gullu-Tekin, Dengiz:2013hka} and $\hat{G}({\bf x}, {\bf x}^{'})$ denotes the time-integrated scalar Green's function defined as  
\begin{equation}
  \hat{G}({\bf x},{\bf x}^{'})=\int d t^{'} \, G ({\bf x},{\bf x}^{'},t,t^{'})=\frac{1}{4\pi r} \mbox{erf} (\frac{Mr}{2}).
\end{equation}
Finally, the Newtonian potential energy can be obtained as
\begin{equation}
 \begin{aligned}
{\cal U}=&-\frac{Gm_1m_2}{r}\mbox{erf} (\frac{Mr}{2})+\frac{M^3}{2\sqrt{\pi}}e^{-\frac{M^2r^2}{4}}G[J_1.J_2-(J_1.\hat{r})(J_2.\hat{r})]\\&-G[J_1.J_2-3(J_1.\hat{r})(J_2.\hat{r})]\times\bigg[\frac{1}{r^3}\mbox{erf} (\frac{Mr}{2})-\frac{M}{\sqrt{\pi}r^2}e^{-\frac{M^2r^2}{4}}\bigg].
  \label{newtpot1}
  \end{aligned}
\end{equation}
Observe that the first term is the ordinary potential energy in IDG which was found in \cite{Biswas1}, and the last two terms are the spin-spin part which could be attractive or repulsive depending on the choice of spin alignments. Let us now turn our attention to the small and large distance behaviors of potential energy. For the large separations as $r\to \infty$, $\mbox{erf}(r)\to 1$, $e^{-r^2}\to 0$, then potential energy takes the form
 \begin{equation}
 \begin{aligned}
{\cal U}=-\frac{Gm_1m_2}{r}-\frac{G}{r^3}\bigg( J_1.J_2-3(J_1.\hat{r})(J_2.\hat{r})\bigg),
  \label{newtpot11}
  \end{aligned}
\end{equation}
which reproduces the pure GR result \cite{Gullu-Tekin} as expected. That is, the first term is the usual Newtonian potential energy, and the second one is the spin-spin interactions in GR. On the other side, for the small distances, as expanding the error and the exponential functions into series around $r=0$ give 
 \begin{equation}
\mbox{erf}(r)= \frac{2r}{\sqrt{\pi}}-\frac{2r^3}{3\sqrt{\pi}}+ {\cal{O}}(r^5),\hskip .6 cm e^{-r^2}=1-r^2+{\cal{O}}(r^4),
 \end{equation}
 the potential energy reads
\begin{equation}
 \begin{aligned}
{\cal U}=&-\frac{ G m_1m_2 M}{\sqrt{\pi}}+\frac{GM^3}{3\sqrt{\pi}}J_1.J_2+{\cal{O}}(r^2).
  \label{newtpot12}
  \end{aligned}
\end{equation}
Here, the ordinary Newtonian potential term and the spin-spin interaction term  in (\ref{newtpot12}) are constant and hence the potential is not singular at the origin. In GR, the spin-spin part diverges according to $\sim -\frac{1}{r^3}$ \cite{Gullu-Tekin}, whereas in the IDG, this part is non-singular. Though the potential energy is generated by matter sources which have dirac delta function singularities, it is regular due to the non-locality. Thus, in the IDG, not only the usual Newtonian potential but also the spin-spin part become regular as one approaches $r\to 0$. Therefore, the theory has improved behavior in the small scale behavior.

\section{Further Gravitomagnetism effects in IDG}
In the previous part, we have shown that both usual Newtonian potential and spin-spin terms are finite at the origin. This is a remarkable result, but one can ask whether further gravitomagnetic effects such as spin-orbit interactions also have non-singular behavior or not. To answer this question, let us turn our attention to the tree-level scattering amplitude between two spinning sources that also have velocities and orbital motion. For this purpose, let us consider the following energy-momentum tensors \cite{Weinberg}:
\begin{eqnarray} 
	T_{00}=T^{\left(0\right)}_{00}+T^{\left(2\right)}_{00}, \hspace{10 mm}
	T_{i0}=T^{\left(1\right)}_{i0}, \hspace{10 mm}
	T_{ij}=T^{\left(2\right)}_{ij} , 
	\label{Source_mom}
\end{eqnarray}
where the relevant tensors are
\begin{eqnarray} 
	T^{\left(0\right)}_{00}&=&m_a\delta^{(3)}\left(\vec{x}-\vec{x}_{a}\right),\nonumber \\
	T^{\left(2\right)}_{00}&=&\frac{1}{2}m_a\vec{v}_a^{2} \delta^{(3)} \left(\vec{x}-\vec{x}_{a}\right)-\frac{1}{2}J_a^{k}\,v_a^{i}\epsilon^{ikj}\partial_{j}\delta^{(3)} \left(\vec{x}
	-\vec{x}_{a}\right),\nonumber \\
	T^{\left(1\right)}_{i0}&=&-m_a v_a^{i} \delta^{(3)} \left(\vec{x}-\vec{x}_{a}\right)+\frac{1}{2}J_a^{k}\,\epsilon^{ikj}\partial_{j}\delta^{(3)}\left(\vec{x}
	-\vec{x}_{a}\right),\nonumber \\
	T^{\left(2\right)}_{ij}&=&m_av_a^{i}v_a^{j}\delta^{(3)} \left(\vec{x}-\vec{x}_{a}\right)+J_a^{l}v_a^{(i}\epsilon^{j)kl}\partial_{k}\delta^{(3)} \left(\vec{x}
	-\vec{x}_{a}\right). 
	\label{en_mom}
\end{eqnarray}
Here, $\vec{v}_{i}$ are the velocities of the particles as defined in a rest frame, and $v^{(i}\epsilon^{j)kl}$ denotes symmetrization.  We shall work in the small velocity and spin limits, in other words up to $O(v^{2})$ and $O(vJ)$. In this respect, the scattering amplitude (\ref{mainressct}) turns into

\begin{equation}
	4A=-2\kappa T{}_{00}^{\prime}( a(\Box)\partial^{2})^{-1}T^{00} -4\kappa T{}_{0i}^{\prime}( a(\Box)\partial^{2})^{-1}T^{0i} 
	-2\kappa T{}_{ij}^{\prime}( a(\Box)\partial^{2})^{-1}T^{ij}+\kappa  T^{\prime}( a(\Box)\partial^{2})^{-1}T,
	\label{sctIDG}
\end{equation}
where integral signs are suppressed and $( a(\Box)\partial^{2})^{-1}$ is the scalar Green's function as was given in (\ref{decgreen1}). To find the weak field potential energy for the sources given in (\ref{Source_mom}), let us calculate the amplitude by working each term in (\ref{sctIDG}), separately. After evaluating the relevant integrals, the energy density interaction term takes the form
\begin{equation}
\begin{aligned}
	-2\kappa T{}_{00}( a(\Box)\partial^{2})^{-1}T^{\prime 00}&=-2\kappa \bigg [ \frac{m_{1}m_{2}}{4\pi r}\bigg (1+\frac{\vec{v}^{2}_{1}+\vec{v}^{2}_{2}}{2}\bigg )\mbox{erf} (\frac{Mr}{2})\\&+\frac{1}{4\pi}\bigg(\frac{1}{ r^{2}}\mbox{erf} (\frac{Mr}{2})-\frac{M}{\sqrt{\pi} r}e^{-\frac{M^2r^2}{4}}\bigg)
	\bigg (\frac{m_{1}(\hat{r}\times \vec{v}_{2})\cdot\vec{J_{2}}}{2}-\frac{m_{2}(\hat{r}\times \vec{v}_{1} )\cdot\vec{J_{1}}}{2} \bigg )\bigg ]t.
\end{aligned}
\end{equation}

Here, we have dropped the term which includes higher order contributions $O(J^{2}v^{2})$. On the other hand, the trace-trace interaction term yields 
\begin{equation}
\begin{aligned}
	\kappa T^{\prime}( a(\Box)\partial^{2})^{-1}T=&	\kappa\bigg [\frac{m_{1}m_{2}}{4\pi r}\bigg(1+\frac{-\vec{v}^{2}_{1}-\vec{v}^{2}_{2}}{2}\bigg)\mbox{erf} (\frac{Mr}{2})\\&
	+\frac{1}{4\pi}\bigg(\frac{1}{ r^{2}}\mbox{erf} (\frac{Mr}{2})-\frac{M}{\sqrt{\pi} r}e^{-\frac{M^2r^2}{4}}\bigg)\bigg (
	-\frac{m_{1}(\hat{r}\times \vec{v}_{2})\cdot\vec{J_{2}}}{2}+
	\frac{m_{2}(\hat{r}\times \vec{v}_{1})\cdot\vec{J_{1}}}{2} \bigg )\bigg ] t.
\end{aligned}
\end{equation}
Similarly the $T{}_{0i}^{\prime}(\partial^{2})^{-1}T^{0i}$ term becomes 
\begin{equation}
\begin{aligned}
-4\kappa T{}_{0i}^{\prime}( a(\Box)\partial^{2})^{-1}T^{0i}&=-4\kappa \bigg [ -\dfrac{m_{1}m_{2}\vec{v}_{1}\cdot\vec{v}_{2} }{4\pi r}\mbox{erf} (\frac{Mr}{2}) \\&+\frac{1}{8\pi}\bigg(\dfrac{1}{r^{2}} \mbox{erf} (\frac{Mr}{2})-\frac{M}{\sqrt{\pi}r}
e^{-\frac{M^2r^2}{4}}\bigg)   \bigg ( -m_{1}(\hat{r}\times \vec{v}_{1}  )\cdot \vec{J}_{2}+m_{2}(\hat{r}\times \vec{v}_{2}  )\cdot \vec{J}_{1}\bigg)\\&  - \dfrac{1}{16 \pi }\bigg( \frac{M^3}{2\sqrt{\pi}}e^{-\frac{M^2r^2}{4}}[J_1.J_2-(J_1.\hat{r})(J_2.\hat{r})]\\&-[J_1.J_2-3(J_1.\hat{r})(J_2.\hat{r})]\times[\frac{1}{r^3}\mbox{erf} (\frac{Mr}{2})-\frac{M}{\sqrt{\pi}r^2}e^{-\frac{M^2r^2}{4}}]\bigg)\bigg]t.
\end{aligned}
\end{equation}
Note that as the $T{}_{ij}^{\prime}(\partial^{2})^{-1}T{}^{ij}$  term in (\ref{sctIDG}) contributes only at the higher order, it has been dropped.
Consequently, by using all these results, the potential energy in IDG takes the form

\begin{equation}
\begin{aligned}
U_{IDG} =& -\frac{G}{r}  m_{1}m_{2} \left [ 1+\frac{3}{2}\vec{v}^{2}_{1}+\frac{3}{2}\vec{v}^{2}_{2}-4\vec{v}_{1}\cdot \vec{v}_{2} \right ] \mbox{erf} (\frac{Mr}{2})
+\frac{M^3}{2\sqrt{\pi}}e^{-\frac{M^2r^2}{4}}G[J_1.J_2-(J_1.\hat{r})(J_2.\hat{r})]\\&-G[J_1.J_2-3(J_1.\hat{r})(J_2.\hat{r})]\times\bigg[\frac{1}{r^3}\mbox{erf} (\frac{Mr}{2})-\frac{M}{\sqrt{\pi}r^2}e^{-\frac{M^2r^2}{4}}\bigg]\\
&-G\bigg(\frac{1}{r^{2}}\mbox{erf} (\frac{Mr}{2})-\frac{M}{\sqrt{\pi}r} e^{-\frac{M^2r^2}{4}}\bigg) \bigg [ \frac{3m_{1}(\hat{r}\times \vec{v}_{2})\cdot\vec{J_{2}}}{2}-\frac{3m_{2}(\hat{r}\times \vec{v}_{1})\cdot\vec{J_{1}}}{2}\\&-2m_{1}(\hat{r}\times \vec{v}_{1})\cdot\vec{J_{2}}+2m_{2}(\hat{r}\times \vec{v}_{2})\cdot\vec{J_{1}}
	\bigg ]. \label{IDGgrpe}
	\end{aligned}
\end{equation}
Observe that potential energy has the ordinary Newtonian potential energy, spin-spin and spin-orbit interactions. 
For large separations as $r\to \infty$, the potential energy becomes
 \begin{equation}
 \begin{aligned}
U =& -\frac{G}{r}  m_{1}m_{2} \left [ 1+\frac{3}{2}\vec{v}^{2}_{1}+\frac{3}{2}\vec{v}^{2}_{2}-4\vec{v}_{1}\cdot \vec{v}_{2} \right ]
-\frac{G}{r^{3}}\left[\vec{J_{1}}\centerdot\vec{J_{2}}-3\vec{J_{1}}\centerdot\hat{r}\,
\vec{J_{2}}\centerdot\hat{r}\right] \\
&-\frac{G}{r^{2}}  \left [ \frac{3m_{1}(\hat{r}\times \vec{v}_{2})\cdot\vec{J_{2}}}{2}-\frac{3m_{2}(\hat{r}\times \vec{v}_{1})\cdot\vec{J_{1}}}{2}-2m_{1}(\hat{r}\times \vec{v}_{1})\cdot\vec{J_{2}}+2m_{2}(\hat{r}\times \vec{v}_{2})\cdot\vec{J_{1}}
	\right ], 
\end{aligned}	
 \end{equation}
which matches with the pure GR result \cite{Tasseten} as expected. That is, the potential energy contains the usual Newtonian potential energy and relativistic corrections. On the other hand, for small distances, the potential energy reduces to
\begin{equation}
 \begin{aligned}
{\cal U}=&-\frac{ G m_1m_2 M}{\sqrt{\pi}}\bigg[ 1+\frac{3}{2}\vec{v}^{2}_{1}+\frac{3}{2}\vec{v}^{2}_{2}-4\vec{v}_{1}\cdot \vec{v}_{2} \bigg]+\frac{GM^3}{3\sqrt{\pi}}J_1.J_2+{\cal{O}}(r).
  \label{newtpot121}
  \end{aligned}
\end{equation}
Here, the ordinary Newtonian potential term and the spin-spin interaction term  in (\ref{newtpot121}) are constant and the spin-orbit interaction terms contribute at the order ${\cal{O}}(r)$. Therefore the potential is regular at the origin. Thus, in the IDG, not only the usual Newtonian potential but also the spin-spin and spin-orbit interactions become regular as one approaches $r\to 0$. These non-singular results in IDG show that the theory is very well-behaved in the UV region compared to GR.

\section{Conclusions and further discussions}
We have considered the IDG in $3+1$ dimensional flat backgrounds. We computed the tree-level scattering amplitude in IDG and accordingly found weak field potential energy between two point-like spinning sources interacting via one-graviton exchange. We have demonstrated that at large distances potential energy is the same as the GR result, whereas at small distances, it is discreetly different from GR. We have also shown that both the ordinary Newtonian potential energy  and the spin-spin term remain finite at the small distance limit ($r\to 0$). Furthermore, in addition to spin-spin interactions, we studied the spin-orbit interactions in IDG by considering that the sources are also moving. We found that not only mass-mass but also spin-spin and spin-orbit interactions are non-singular and finite at the origin. That is, gravitational potential energy of spinning sources that also have velocities becomes non-singular for IDG. Consequently, the theory is a very well-behaved feature in the UV regime as compared to GR. 

Now, we would like to discuss the effects of mass scale of non-locality ($M$) on gravitational memory effect. Gravitational waves, induced by merger of neutron stars or black holes etc, create a  permanent effect on a system composed of inertial test particles. In other words, a pulse of gravitational wave changes the relative displacements of test particles. This effect is called gravitational memory effect and comes in two forms: ordinary (or linear) \cite{zeldovich} and null  (or
non-linear) \cite{Christodoulou}. The studies on gravitational memory effect have recently received more attention in various aspects \cite{Garfinkle,Satishchandran:2017pek,Tolish1,Tolish2,Bieri2,Gibbons,Kilicarslan} because there is a hope that it could be measured by advanced LIGO.  To calculate gravitational memory effect in IDG in a flat spacetime, we can follow the method of \cite{Satishchandran:2017pek, Garfinkle}: we first solved the geodesic deviation equation and then integrated it two times to find relative separation of the test particles.   Without giving the details, we shall  give the final result:
\begin{equation}
\begin{aligned}
\Delta\xi^{i}&
=\frac{1}{r} \mbox{erf}(\frac{Mr}{2})\Delta_j^i\Theta(U)\xi^{j},
\label{memoryeffect}
\end{aligned}
\end{equation} 
where $\Theta$ is the step function, $\xi$ is a spatial separation vector and $\Delta_j^i$ are spatial components of the memory tensor (See Eq.(45) in \cite{Garfinkle} for memory tensor). This result shows that the test particles have non-trivial change in their separations which is described by the memory tensor. Observe that the memory is dependent of the mass scale of non-locality and different from GR. In the large distance limits, memory is the same as the usual Einsteinian form as expected. Furthermore, for a lower bound on mass scale of non-locality ($M>4keV$) \cite{Edholm:2018qkc}, the memory reduces to GR prediction above at very small distances.

\section{\label{ackno} Acknowledgements}
We would like to thank B. Tekin for useful discussions, suggestions and comments. We would also like to thank S.Dengiz and J. Edholm  for suggestions and critical readings of the manuscript.

\section{Appendix: Details of the Calculations}
In this part, we would like to give the details of scattering amplitude calculations for the Sec. III. Before going into further details,  let us give the following identities:
\begin{align}
\partial_{k}r= \dfrac{(x^{k}-x^{\prime k})}{r}=\hat{r}^{k}, \hspace{1cm} \partial_{k}\dfrac{1}{r}=\dfrac{-(x^{k}-x^{\prime k})}{r^{3}}=\dfrac{-\hat{r}^{k}}{r^{2}}, \nonumber \\
\partial_{k^{\prime}}r= \dfrac{-(x^{k}-x^{\prime k})}{r}=-\hat{r}^{k}, \hspace{1cm} \partial_{k^{\prime}}\dfrac{1}{r}=\dfrac{(x^{k}-x^{\prime k})}{r^{3}}=\dfrac{\hat{r}^{k}}{r^{2}}, \nonumber \\
\partial_{k}\partial_{n^{\prime}}r=\dfrac{1}{r}\left( -\delta^{kn}+\hat{r}^{k}\hat{r}^{n}\right) , \hspace{1cm}
\partial_{k}\partial_{n^{\prime}}\dfrac{1}{r}= \dfrac{1}{r^{3}}\left( \delta^{kn}-3\hat{r}^{k}\hat{r}^{n}\right),\nonumber \\
\partial_{k}\mbox{erf} (r)=\frac{2}{\sqrt{\pi}}e^{-r^2}\hat{r}^{k} , \hspace{1cm}
\partial_{k^{\prime}}\mbox{erf} (r)=-\frac{2}{\sqrt{\pi}}e^{-r^2}\hat{r}^{k},
\end{align}
which are needed for computations. Let us now calculate the amplitude by working each term in (\ref{sctIDG}), separately. The energy density interaction term becomes

\begin{equation}
\begin{aligned}
	T{}_{00}(a(\Box)\partial^{2})^{-1}T^{\prime 00}=&\bigg [  m_{1}\delta^{(3)}\left(\vec{x}-\vec{x}_{1}\right)+\frac{1}{2}m_{1}\vec{v}^{2}_{1} \delta^{(3)}\left(\vec{x}-\vec{x}_{1}\right) -\frac{1}{2}J_{1}^{l}\,v^{i}_{1}\epsilon^{ilk}\partial_{k}\delta\left(\vec{x}
	-\vec{x}_{1}\right)\bigg ]  (a(\Box)\partial^{2})^{-1} \\
	&
	\bigg [ m_{2}\delta^{(3)}\left(\vec{x^{\prime}}-\vec{x}_{2}\right)+\frac{1}{2}m_{2}\vec{v}^{2}_{2} \delta^{(3)}\left(\vec{x^{\prime}}-\vec{x}_{2}\right)  -\frac{1}{2}J_{2}^{m}\,v^{j}_{2}\epsilon^{jmn}\partial_{n}^{\prime}\delta^{(3)}\left(\vec{x^{\prime}}
	-\vec{x}_{2}\right)\bigg ],
	\end{aligned}
\end{equation}
whose each distinct term reads
\begin{equation}
	m_{1}\delta^{(3)}\left(\vec{x}-\vec{x}_{1}\right)( a(\Box)\partial^{2})^{-1}m_{2}\delta^{(3)}\left(\vec{x^{\prime}}-\vec{x}_{2}\right)=\frac{m_{1}m_{2}}{4\pi r}\mbox{erf} (\frac{Mr}{2})t,
\end{equation}

\begin{equation}
	m_{1}\delta^{(3)}\left(\vec{x}-\vec{x}_{1}\right)( a(\Box)\partial^{2})^{-1}\frac{1}{2}m_{2}\vec{v}^{2}_{2} \delta^{(3)}\left(\vec{x^{\prime}}-\vec{x}_{2}\right)=\frac{1}{2}\frac{m_{1}m_{2}\vec{v}^{2}_{2}}{4\pi r}\mbox{erf} (\frac{Mr}{2})t,
\end{equation}

\begin{equation}
\begin{aligned}
	-\frac{1}{2}m_{1}\delta^{(3)}\left(\vec{x}-\vec{x}_{1}\right)( a(\Box)\partial^{2})^{-1}J_{2}^{m}\,v^{j}_{2}\epsilon^{jmn}\partial_{n}^{\prime}\delta^{(3)}\left(\vec{x^{\prime}}
	-\vec{x}_{2}\right) 
	=&\frac{1}{2}
	\frac{m_{1}(\hat{r}\times \vec{v}_{2}).\vec{J_{2}}}{4\pi r^{2}}\mbox{erf} (\frac{Mr}{2})t\\&-\frac{M}{2\sqrt{\pi}}e^{-\frac{M^2r^2}{4}}
	\frac{m_{1}(\hat{r}\times \vec{v}_{2}).\vec{J_{2}}}{4\pi r}t,	
	\end{aligned}
\end{equation}

\begin{equation}
\frac{1}{2}m_{1}\vec{v}^{2}_{1} \delta^{(3)}\left(\vec{x}-\vec{x}_{1}\right)( a(\Box)\partial^{2})^{-1}m_{2}\delta^{(3)}\left(\vec{x^{\prime}}-\vec{x}_{2}\right)=\frac{1}{2}\frac{m_{1}m_{2}\vec{v}^{2}_{1}}{4\pi r}\mbox{erf} (\frac{Mr}{2})t,
\end{equation}

\begin{equation}
\begin{aligned}
	-\frac{1}{2}J_{1}^{l}\,v^{i}_{1}\epsilon^{ilk}\partial_{k}\delta^{(3)}\left(\vec{x}
	-\vec{x}_{1}\right) ( a(\Box)\partial^{2})^{-1} m_{2}\delta^{(3)}\left(\vec{x^{\prime}}-\vec{x}_{2}\right)
	=&-\frac{1}{2}
	\frac{m_{2}(\hat{r}\times \vec{v}_{1}).\vec{J_{1}}}{4\pi r^{2}}\mbox{erf} (\frac{Mr}{2})t\\&+\frac{M}{2\sqrt{\pi}}
	\frac{m_{2}(\hat{r}\times \vec{v}_{1}).\vec{J_{1}}}{4\pi r}e^{-\frac{M^2r^2}{4}}t,
	\end{aligned}
\end{equation}
with these terms, one ultimately gets
\begin{equation}
\begin{aligned}
	-2\kappa T{}_{00}( a(\Box)\partial^{2})^{-1}T^{\prime 00}&=-2\kappa \bigg [ \frac{m_{1}m_{2}}{4\pi r}\mbox{erf} (\frac{Mr}{2})\bigg (1+\frac{\vec{v}^{2}_{1}+\vec{v}^{2}_{2}}{2}\bigg )\\&+\frac{1}{4\pi}\bigg(\frac{1}{r^{2}}\mbox{erf} (\frac{Mr}{2})-\frac{M}{\sqrt{\pi} r}e^{-\frac{M^2r^2}{4}}\bigg)
	\bigg (\frac{m_{1}(\hat{r}\times \vec{v}_{2})\cdot\vec{J_{2}}}{2}-\frac{m_{2}(\hat{r}\times \vec{v}_{1} )\cdot\vec{J_{1}}}{2} \bigg )\bigg ]t.
\end{aligned}
\end{equation}

On the other side, the trace-trace interaction term yields 
\begin{equation}
\begin{aligned}
	T^{\prime}( a(\Box)\partial^{2})^{-1}T= &\bigg [ -m_{1}\delta^{(3)}\left(\vec{x}-\vec{x}_{1}\right)+\frac{1}{2}m_{1}\vec{v}^{2}_{1} \delta^{(3)}\left(\vec{x}-\vec{x}_{1}\right) -\frac{1}{2}J_{1}^{l}\,v^{i}_{1}\epsilon^{ilk}\partial_{k}\delta^{(3)}\left(\vec{x}
	-\vec{x}_{1}\right)\bigg ] (\partial^{2})^{-1} \\
	&\bigg [ -m_{2}\delta^{(3)}\left(\vec{x^{\prime}}-\vec{x}_{2}\right)+\frac{1}{2}m_{2}\vec{v}^{2}_{2} \delta^{(3)}\left(\vec{x^{\prime}}-\vec{x}_{2}\right)  -\frac{1}{2}J_{2}^{m}\,v^{j}_{2}\epsilon^{jmn}\partial_{n}^{\prime}\delta^{(3)}\left(\vec{x^{\prime}}
	-\vec{x}_{2}\right)\bigg ].
\end{aligned}
\end{equation}
Then, by evaluating the relevant integrals, one eventually obtains 
\begin{equation}
\begin{aligned}
	\kappa T^{\prime}( a(\Box)\partial^{2})^{-1}T=&	\kappa\bigg [\frac{m_{1}m_{2}}{4\pi r}\bigg(1+\frac{-\vec{v}^{2}_{1}-\vec{v}^{2}_{2}}{2}\bigg)\mbox{erf} (\frac{Mr}{2})\\&
	+\bigg(\frac{1}{4\pi r^{2}}\mbox{erf} (\frac{Mr}{2})-\frac{M}{4\pi^{\frac{3}{2}} r}e^{-\frac{M^2r^2}{4}}\bigg)\bigg (
	-\frac{m_{1}(\hat{r}\times \vec{v}_{2})\cdot\vec{J_{2}}}{2}+
	\frac{m_{2}(\hat{r}\times \vec{v}_{1})\cdot\vec{J_{1}}}{2} \bigg )\bigg ] t.
\end{aligned}
\end{equation}
Similarly, the $T{}_{0i}^{\prime}(\partial^{2})^{-1}T^{0i}$ term can be written as
\begin{equation}
\begin{aligned}
	T{}_{0i}^{\prime}( a(\Box)\partial^{2})^{-1}T^{0i}=&\bigg [ -m_{1}v^{i}_{1}\delta^{(3)}\left(\vec{x}-\vec{x}_{1}\right)+\frac{1}{2}J_{1}^{k}\,\epsilon^{ikj}\partial_{j}\delta^{(3)}\left(\vec{x}
	-\vec{x}_{1}\right)\bigg ](a(\Box)\partial^{2})^{-1}\\&\times\bigg [ m_{2}v^{i}_{2}\delta^{(3)}\left(\vec{x^{\prime}}-\vec{x}_{2}\right)-\frac{1}{2}J_{2}^{l}\,\epsilon^{ilm}\partial^{\prime}_{m}\delta^{(3)}\left(\vec{x^{\prime}}
	-\vec{x}_{2}\right)\bigg ],
\end{aligned}
\end{equation}
which after lengthy and tedious calculations becomes
\begin{equation}
\begin{aligned}
-4\kappa T{}_{0i}^{\prime}( a(\Box)\partial^{2})^{-1}T^{0i}&=-4\kappa \bigg [ -\dfrac{m_{1}m_{2}\vec{v}_{1}\cdot\vec{v}_{2} }{4\pi r}\mbox{erf} (\frac{Mr}{2}) \\&+\frac{1}{8\pi}\bigg(\dfrac{1}{ r^{2}} \mbox{erf} (\frac{Mr}{2})-\frac{M}{\sqrt{\pi} r}
e^{-\frac{M^2r^2}{4}}\bigg)  \bigg ( -m_{1}(\hat{r}\times \vec{v}_{1}  )\cdot \vec{J}_{2}+m_{2}(\hat{r}\times \vec{v}_{2}  )\cdot \vec{J}_{1}\bigg)\\&  - \dfrac{1}{16 \pi }\bigg( \frac{M^3}{2\sqrt{\pi}}e^{-\frac{M^2r^2}{4}}[J_1.J_2-(J_1.\hat{r})(J_2.\hat{r})]\\&-[J_1.J_2-3(J_1.\hat{r})(J_2.\hat{r})]\times(\frac{1}{r^3}\mbox{erf} (\frac{Mr}{2})-\frac{M}{\sqrt{\pi}r^2}e^{-\frac{M^2r^2}{4}})\bigg)\bigg]t.
\end{aligned}
\end{equation}
Recall that the $T{}_{ij}^{\prime}(\partial^{2})^{-1}T{}^{ij}$  term contributes in higher order corrections.
Consequently, by using the results above obtained, the potential energy in IDG is obtained in the form as given in (\ref{IDGgrpe}).

\end{document}